\documentclass[sigconf]{acmart}

\AtBeginDocument{%
  \providecommand\BibTeX{{%
    \normalfont B\kern-0.5em{\scshape i\kern-0.25em b}\kern-0.8em\TeX}}}

\usepackage{subfigure}
\usepackage{enumitem}
\usepackage{makecell}
\usepackage{longtable,booktabs,etoolbox}
\usepackage{array,booktabs,enumitem}
\newcolumntype{P}[1]{>{\endgraf\vspace*{-\baselineskip}}p{#1}}
\usepackage{multirow}
\usepackage{enumitem}
\usepackage{dblfloatfix} 
\usepackage{arydshln}
\usepackage{titlesec}

\copyrightyear{2021} 
\acmYear{2021} 
\setcopyright{acmcopyright}\acmConference[HAI '21]{Proceedings of the 9th International Conference on Human-Agent Interaction}{November 9--11, 2021}{Nagoya, Japan}
\acmBooktitle{Proceedings of the 9th International Conference on Human-Agent Interaction (HAI '21), November 9--11, 2021, Nagoya, Japan}
\acmPrice{15.00}
\acmDOI{10.1145/3472307.3484182}
\acmISBN{978-1-4503-8620-3/21/11}

 \usepackage[pscoord]{eso-pic}
\newcommand{\placetextbox}[3]{
 \setbox0=\hbox{#3}
 \AddToShipoutPictureFG*{ \put(\LenToUnit{#1\paperwidth},\LenToUnit{#2\paperheight}){\vtop{{\null}\makebox[0pt][c]{#3}}}
 }
 }
  \placetextbox{.5}{0.055}{\footnotesize{© {ACM} {2021}. This is the author's version of the work. It is posted here for your personal use. Not for redistribution.}}
  \placetextbox{.5}{0.045}{\footnotesize{The definitive Version of Record was published in HAI '21: Proceedings of the 9th International Conference on Human-Agent Interaction, https://doi.org/10.1145/3472307.3484182.}}

\begin{document}

\title{What are Social Norms for Low-speed Autonomous Vehicle Navigation in Crowded Environments? An Online Survey}

\author{Mahsa Golchoubian}
\affiliation{%
  \institution{Department of Systems Design Engineering, \\University of Waterloo}
  \city{Waterloo}
  \country{Canada}}
\email{mahsa.golchoubian@uwaterloo.ca}

\author{Moojan Ghafurian}
\affiliation{%
  \institution{Department of Electrical and Computer Engineering, University of Waterloo}
  \city{Waterloo}
  \country{Canada}}
\email{moojan@uwaterloo.ca}

\author{Nasser Lashgarian Azad}
\affiliation{%
  \institution{Department of Systems Design Engineering, \\University of Waterloo}
  \city{Waterloo}
  \country{Canada}}
\email{nlashgarianazad@uwaterloo.ca}

\author{Kerstin Dautenhahn}
\affiliation{%
  \institution{Departments of Electrical and Computer Engineering and Systems Design Engineering, University of Waterloo}
  \city{Waterloo}
  \country{Canada}}
\email{kerstin.dautenhahn@uwaterloo.ca}

\renewcommand{\shortauthors}{Golchoubian, et al.}

\begin{abstract}
  It has been suggested that autonomous vehicles can improve efficiency and safety of the transportation systems. While research in this area often focuses on autonomous vehicles which operate on roads, the deployment of low-speed, autonomous vehicles in unstructured, crowded environments has been studied less well and requires specific considerations regarding their interaction with pedestrians. For making the operation of these vehicles acceptable, their behaviour needs to be perceived as safe by both pedestrians and the passengers riding the vehicle. In this paper we conducted an online survey with 116 participants, to  understand people’s preferences with respect to an autonomous golf cart’s behaviour in different interaction scenarios. We measured people’s self-reported perceived safety towards different behaviour of the cart in a variety of scenarios. Results suggested that despite the unstructured nature of the environment, the cart was expected to follow common traffic rules when interacting with a group of pedestrians.
\end{abstract}




\maketitle

\section{Introduction}
For many years researchers have studied the automation of  road-based transportation systems and  proposed different algorithms for navigation and decision making of autonomous vehicles~\cite{schwarting2018planning,milani2020smart}. Recently, there has been an increasing emphasis on studying pedestrians' interaction with these autonomous vehicles (AV) in crossing zones and ways the AVs can communicate with pedestrians in the absence of a driver~\cite{rasouli2019autonomous}. However, to the best of our knowledge, the existing studies have considered the road structure as the operational environment. It is not clear if the same results would be obtained if we want to operate smaller, lower speed autonomous vehicles in unstructured, crowded environments such as airport terminals or shopping malls.

At present, we can find golf carts or so-called buggies in airport terminals or big shopping malls that are being used for transporting people. Unlike road structures where the space is dominated by the vehicles, in these indoor environments the space is occupied mostly by pedestrians and therefore the interaction that the vehicle will have with the pedestrians should be a major focus to make the operation of these vehicles safe and socially acceptable.

The number of golf cart related injuries in the U.S. was reported to be increasing between 1990 and 2006~\cite{watson2008golf}, it is expected that the growth in population may lead to higher figures today. Being ran over by the cart (16.2\%) was reported to be one of the two primary causes of these injuries~\cite{watson2008golf}, which could be prevented by better controlled driving. This number of injuries may  increase even more when these carts are operated in crowded areas. As the operation of these golf carts is different from automobiles, their drivers require special training for a safe operation to account for, e.g., the speed of the cart and distance to the pedestrians. We assume that the use of autonomous carts in these environments, beside cutting off operational cost (e.g. hiring and training a driver), could bring more safety and comfort to the operation of these carts.

Along with increasing safety by reducing the chance of hitting a pedestrian (e.g., through obstacle recognition and avoidance methods), the operation of these autonomous carts also needs to be {\it perceived} safe by pedestrians and passengers. Therefore, it is important to understand people’s expectations of the cart’s behaviour while moving  in a crowd, which can identify social norms that the cart should follow to make it  more acceptable.

Therefore, to inform our future steps of developing a socially compliant navigation algorithm for the operation of these autonomous, low speed vehicles in pedestrian-rich unstructured environments, we conducted an online study to understand people's preferences about the path of a small autonomous car operating among pedestrians. We used sketched interaction scenarios and asked participants about their opinions and feelings towards different possible behaviours of the autonomous golf cart in each of the scenarios.

\section{Background}

Sociability is an important factor influencing a robot’s acceptance~\cite{de2013exploring}, e.g., robots can gain people's trust in a navigation task~\cite{rossi2019investigating}. In the same way human-like driving behaviour of autonomous vehicles can improve acceptance and trust~\cite{oliveira2019driving}. For a navigation process to be considered socially compliant, three main aspects of comfort, naturalness, and sociability should be satisfied~\cite{kruse2013human}. In the following, an overview of these three aspects and the related existing work are discussed.

\textit{Comfort:} The effect that respecting humans’ personal space, known as proxemics, has on people’s comfort has been studied widely in the literature (e.g., see~\cite{rios2015proxemics,takayama2009influences,lauckner2014hey}). Therefore, many researchers have paid attention to the aspect of comfort by ensuring that the robot maintains an appropriate distance from each person during its navigation in a crowd~\cite{zhang2020motion,rios2011understanding,rios2012navigating}. Mutually respected spatial distances for human-human interaction in static positions has been found in~\cite{hall1968proxemics} and the same distances can be used for a human-sized robot as well. Pedestrian’s tracking data has shown that humans usually keep a similar minimum distance when passing each other in a pairwise interaction~\cite{corbetta2018physics}. The appropriate social distancing for circumventing a human was studied in~\cite{herrera2019cognitive} for a small Pioneer 3-AT mobile robot. Note, these social distances depend on the robot or the vehicle’s size as well as its speed. Studies have also shown the strong effect of the autonomous vehicles' speed~\cite{schneemann2016analyzing} and distance~\cite{burns2019pedestrian} on pedestrians' crossing decisions, and it has been suggested that these criteria may have a more dominant effect than the vehicle's  size~\cite{dey2017impact}.

\textit{Naturalness}, as one of the other aspects of socially compliant navigation~\cite{kruse2013human}, is attempted by designing the robots' paths to be smooth, i.e. without sharp deviation angles and velocity changes. Specific algorithms have also been used in mimicking human navigation behaviours to achieve the most natural looking paths and collision avoidance techniques for a robot~\cite{fahad2018learning,konar2021learning,tsai2020generative}. Natural movements can increase perceived safety by making the movement more predictable and legible. In autonomous vehicles, often explicit visual and auditory communication means are added for making the vehicle’s movements and intentions clearer and more predictable to the other road users~\cite{dey2020color}.

The \textit{sociability} aspect has not been studied as deeply as the other two aspects. Sociability in robots' navigation can be viewed as robots respecting the social norms in humans’ navigational behaviours. One of the well-known norms in navigation is passing on the right(left) for countries that have right-hand(left-hand) traffic~\cite{zanlungo2012microscopic}. Johnson and Kuipers (2018) designed a navigation algorithm for a robot wheelchair that learned this right-hand norm from observing pedestrians' behaviours~\cite{johnson2018socially}. For inducing the right-hand norm in their socially aware motion planning, Chen et al. (2017) defined an extra term in the reward function of their reinforcement learning algorithm for breaking the symmetry of the left and right passage in favour of the direction compatible with the norm~\cite{chen2017socially}.  Kirby et al. (2009)~\cite{kirby2009companion} coded social conventions such as keeping on the right side of a hallway as a constraint in their navigation algorithm. The social force model was extended in \cite{reddy2020social} by enforcing these social conventions to be followed. These social conventions were also introduced in the strategy selection when using a geometric reactive navigation approach~\cite{reddy2020social}.

Other examples of social norms are when a person excuses oneself and asks for a way through a blocked path. This social behaviour was tested on a robot in \cite{rossi2019investigating} while comparing it to two other non-social behaviours of stopping and waiting for the path to get cleared, and performing a simple obstacle avoidance manoeuvre. 
Empirical observations on humans' behaviour in a crowd has also shown other types of common behaviours such as lane formation in a bidirectional flow area \cite{zhang2019characterization}.

Overall, there are well-known social norms that hold in human-human interaction, and we also have traffic rules that govern human-vehicle and vehicle-vehicle interactions on the roads. But when it comes to the operation of autonomous vehicles in an unstructured environment, which is different from a road situation, it is not clear if and how those conventions will be translated to the different context. As no clear rules like the ones on the road seem to exist in unstructured, crowded environments, behaviours might differ. In the same way, because of the different size and speed of a golf cart, expectations might also differ compared to a small robot interacting with people in the same unstructured environment, i.e., an environment with no specific lane or sign that separates the operational area of these vehicles and where no official regulations exist on how these vehicles should operate in the environment.

To this end, we conducted a study to identify conventions and social norms that people expect from an autonomous cart operating in an unstructured, crowded environment.

\section{Research questions}

Our research questions are as follows:

\textbf{RQ1}: Are there general patterns in the way people (assuming the roles of pedestrians and passengers) expect the cart to behave when it is operating in an unstructured environment like a shopping mall or an airport terminal?

\textbf{RQ2}: For feeling safe, do people (assuming the role of pedestrians) expect the cart to follow rules that are similar to road traffic rules, for passing and overtaking despite the environment being unstructured?

\textbf{RQ3}: How safe and comfortable do people feel when the cart tries to negotiate space with them for passing or crossing?  Are they willing to share their space with these carts and let them pass? What do they think about the priorities in these shared-space environments?

As the nature of this study is exploratory, we did not formulate any hypotheses.

\section{Method}

We designed an online study in which we described and sketched nine common interaction scenarios that could occur between an autonomous golf cart and one or more pedestrians, including scenarios such as passing, overtaking, crossing, turning. The details of each scenario are shown in Table~\ref{tab:scenarios}.

In each interaction scenario, we considered possible behaviours for either the cart or the pedestrian in the sketched scene. The possible designed behaviours for each scenario differed and were limited to the cart’s path selection relative to the crowd, the direction the cart passes or overtakes a pedestrian or other choices such as deceleration, stopping and path changing for collision avoidance, as well as  option of the cart requesting a pass.

We also studied the effect of people's role in the scene on the perceived safety of the cart’s behaviour. We considered two different roles: a pedestrian walking close to the cart, or being a passenger on the cart.

Including people's role and the cart’s behaviour choices, we had a mixed factorial design in each of our scenarios. The participant’s role was a between-subject factor with two levels (pedestrian and passenger), while the cart’s behaviour was a within subject factor with the number of available choices ranging from 2 to 5 depending on the scenario (Table~\ref{tab:scenarios}).

\subsection{Procedure and Measures}
After giving consent and answering demographic questions (i.e., age, gender, country of residence), the participants were randomly assigned to either the role of a pedestrian or a passenger on the cart. The participants were requested to imagine themselves in the scene in the role they were given and answer the questions from that perspective. An instruction page gave each participant a general view about the concept of the passenger-carrying golf cart operating autonomously in a shopping mall or an airport terminal among the crowd along with some real-world images of these golf carts in an indoor environment. The golf cart was described as an electric vehicle with a max speed of 9.5 mph.

\renewcommand{\arraystretch}{1}
\begin{table}
\Description[Designed questions]{The six questions designed for scenario 5, which were responded on a continuous Likert scale.}
\caption{Designed questionnaire for scenario 5 with a continuous Likert scale from strongly disagree to strongly agree.}\label{tab:additioanl-ques}
\small
\resizebox{\columnwidth}{!}{\begin{tabular}{|l|}
\hline
\textbf{Statement}\\
\hline
I feel comfortable with this behaviour of the cart \\
I think this behaviour is dangerous \\
I think this behaviour is appropriate\\
I would not have expected this behaviour\\
I would trust a cart that shows this behaviour\\
The cart did not seem to be aware of me (the pedestrian) when it showed this behaviour\\
\hline
\end{tabular}}
\end{table}

The participants then answered two pre-survey questions about a human-human interaction scenario, asking about their preferred direction for passing or overtaking another person. The participants then went through the 9 designed interaction scenarios one by one. These scenarios are shown in Table~\ref{tab:scenarios}. For scenarios 1, 2, 5, 6, 8, and~9, the participants completed the perceived safety sub-scale of the Godspeed questionnaire for each of the cart’s behaviours. For scenario~9, the subscale of perceived intelligence was also provided. In scenario~5, along with the perceived safety subscale, we also added six other Likert scale questions shown in Table\ref{tab:additioanl-ques}.

Participants evaluated scenarios 3, 4, and 7 in Table~\ref{tab:scenarios} in the form of a multiple-choice questions where the participants responded by choosing all the behavioural options they preferred. In the multiple-choice questions, we also asked participants to explain why they selected a specific option. We used multiple choice questions for these scenarios instead of the Godspeed questionnaire, as we were interested in scenario-specific answers, focusing on the norms related to the operation of the vehicle. 

The survey ended with a set of questions, where the participants were asked about their general concerns towards the operation of these carts in the descried environments, as well as their specific thoughts on the interaction of the cart with children. We included this question in order to probe participants' possible concerns regarding vulnerable populations, in this case restricted to children. We also asked which approach direction may be most concerning when the car is getting close to them.

At the end of the study, the participants saw a \textit{thank you} page and were given an end code to submit their HIT\footnote{Human Intelligence Task (HIT), as called on Amazon Mechanical Turk}.

\begin{figure*}[t!]
\subfigure[scenario 3 - pedestrians]{
\centering \fbox{\includegraphics[height=4cm]{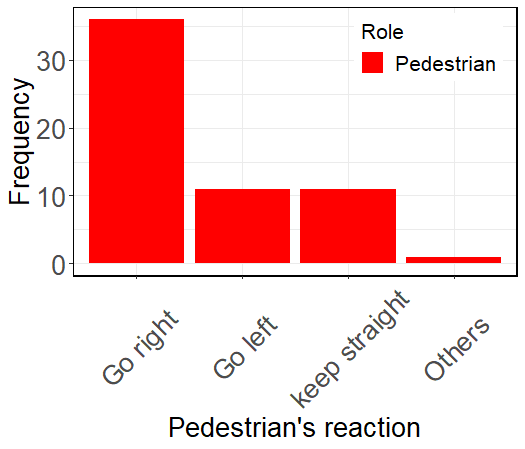}
\Description[A bar graph for pedestrians' preferences in scenario 3]{A bar graph showing the number of times each reaction was chosen by the participants in the role of a pedestrian in scenario 3. The choices from high to low frequencies are: going right, a tie between going left and keeping straight, and others.  }}          
\label{fig:3PED}
}\hspace{2cm} 
\subfigure[scenario 3 - passengers]{
\fbox{\includegraphics[height=4cm]{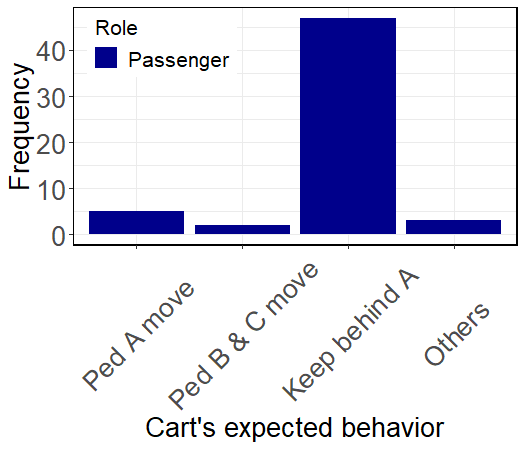}
\Description[A bar graph for passengers' preferences in scenario 3]{A bar graph showing the number of times each behavior for the cart was chosen by the participants in the role of a passenger in scenario 3. The choices from high to low frequencies are: Keeping behind A, pedestrian A moves, others, and pedestrians B \& C move.}}
\label{fig:3PAS}
}\hspace{2cm} \hfill
\subfigure[scenario 4]{
\fbox{\includegraphics[height=4cm]{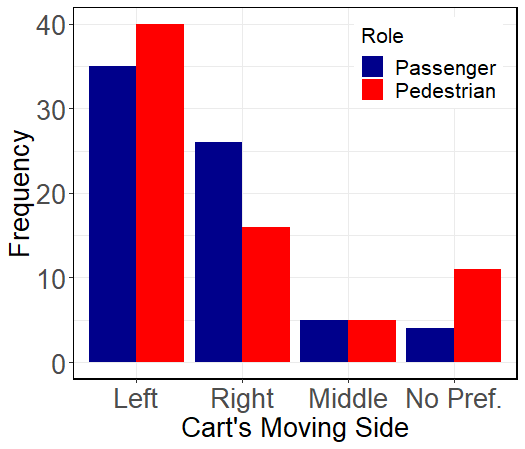}
\Description[A double bar graph for participants' preferences in scenario 4]{A double bar graph showing the number of times each of the left/right/middle/No-preference side of the corridor was chosen for the operation of the cart in scenario 4 by participants in either roles. Pedestrians' choices from high to low frequency are: left, right, no. preference, and middle. Passengers' preferred choices are left, right, middle, and others, respectively}}
\label{fig:s4}
}\hspace{2cm} 
\subfigure[scenario 7]{
\fbox{\includegraphics[height=4cm]{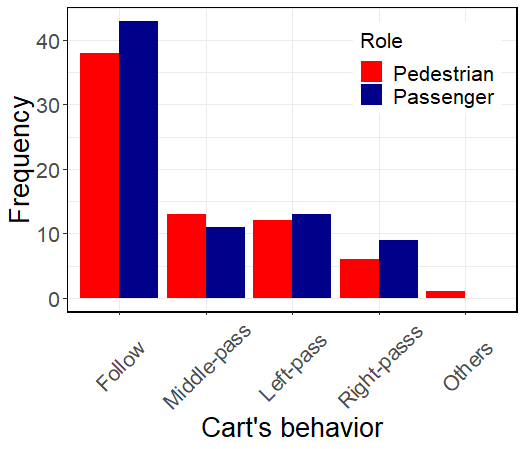}
\Description[A double bar graph for participants' preferences in scenario 7]{A double bar graph showing the number of times each of the cart’s possible behaviors was chosen in scenario 7 by participants in either roles. Pedestrians' choices from high to low frequencies are: follow, middle-pass, left-pass, right-pass, and others. Passengers' choices from high to low frequencies are: follow, left-pass, middle-pass, right-pass, others.}}
\label{fig:s7}
}\hfill
\caption{The results of the multiple choice question scenarios. (a) Pedestrian group's selected reactions about the cart approaching from behind in Scenario 3 (see Table~\ref{tab:scenarios}).
(b) Expectation of the passenger group in the same situation. "Ped" stands for pedestrian. (c) Participants’ preferred side for the cart’s operation in a corridor with an unidirectional flow of pedestrians as in scenario 4. (d) Participants’ expectations of the cart when it gets behind a bidirectional flow of pedestrians as in scenario~7.   }
\label{fig:MultChoice}
\vspace{-12px}
\end{figure*}

\subsection{Participants}
We recruited 131 participants on Amazon Mechanical Turk. The recruitment was limited to US and Canada workers with an approval rate of 95\% or higher based on at least 50 HITs. Fifteen participants failed the attention check question. Therefore, the final data was based on 116 participants' responses (49 Female, 66 Male, 1 Unknown; age range: 18-74). 57 participants were randomly assigned to the role of pedestrian while the other 59 participants were assigned to the role of passenger. The survey took on average 24 minutes to complete and participants were paid $2$ USD upon completing the study and a pro-rated amount if they wished to stop earlier. The study received Ethics clearance from the University of Waterloo Research Ethics Committee.

\subsection{Data Analysis Approach}
 
\textbf{Quantitative data:} Linear mixed-effect models were used for studying significant differences among the perceived safety of the distinct behaviours for the cart in each scenario and between the two different roles. We took potential confounding factors into account and also searched for potential interaction effects in the data. For coming up with one single number for the perceived safety of each behaviour, we averaged the scores of the three questions in the perceived safety subscale of the Godspeed questionnaire while reversing the score of the last question for making it consistent with the other two questions. For multiple choice questions, we performed a Chi-square test to check if there is a significant difference between the selected choices and in case of a significant difference performed pairwise comparisons.

\textbf{Qualitative data:} We conducted thematic analysis on the reasons participants provided for their choices in scenarios 3 and 4. Themes were defined based on the responses and counting the number of times each theme appeared in participants' responses. The results were categorized based on the selected choices.

\section{Results}

\noindent
\begin{table*}
\centering
\small
\Description[Thematic analysis of pedestrians’ reasons in scenario 3]{There are four main reasons for choosing the option of clearing the cart’s path. Three reasons for moving left and four for moving right are also shown. There are two reasons for choosing to keep walking straight without clearing the cart’s path}
\caption{Thematic analysis results on pedestrians' reasons in scenario 3 where the cart gets behind a pedestrian whose left and right side are occupied by other pedestrians and pylons respectively. The table reports the number of times each reason was mentioned. Moving to left/right are sub-choices of clearing the cart's path}\label{tab:s3_PED_Qual}
\begin{tabular}{|l|l|c|l|l|c|}
\hline
\textbf{Choice} & \textbf{Reason} & \textbf{\#} & \textbf{Choice} & \textbf{Reason} & \textbf{\#}\\
\hline
& & & \multirow{3}{3.6cm}{Moving left} & Viewing pylons as a tighter restriction & 5\\
& Uncomfortable/Anxious being followed & 7 & & Feel safer joining other pedestrians & 2\\
\multirow{3}{2.1cm}{Clearing cart's path} & Risk of getting hit & 6 & & Traffic rules (passing on right) & 1
\\\cline{4-6}
& Cart's right of way & 2 & \multirow{4}{1.5cm}{Moving right} & Not disturbing other pedestrians & 14 \\
& Letting faster vehicle overtake & 7 & & Traffic rules (faster vehicle on left) & 5 \\
& & & &  Clear of people & 2 \\
& & & & Intuitive & 2 \\
\hline
\multirow{2}{2cm}{Keep walking straight} & Priority rule & 2 \\
& No room on either side & 7\\
\cline{1-3}
\end{tabular}
\end{table*}

In our single pedestrian-cart passing and overtaking scenarios (scenarios 1 and 2), the passenger group perceived passing on the right much safer than the cart waiting for the pedestrian to clear the path ($se=0.21, t=3.44, p < 0.01$). We did not observe any significant difference in the pedestrian group.

\newcolumntype{C}[1]{>{\centering\arraybackslash}p{#1}}
\newcolumntype{L}{>{\centering\arraybackslash}m{3cm}}

\renewcommand{\arraystretch}{1}
\begin{table}
\centering
\small
\Description[Thematic analysis results for passengers’ reasons in scenario 3]{The eight reasons participants in the role of a passenger mentioned for expecting the cart to stay behind pedestrian A in scenario 3 is shown, along with the number of times each reason was mentioned}
\caption{Thematic analysis results on passengers' reasons for choosing to stay behind person A in scenario 3}\label{tab:s3_PAS_Qual}
\begin{tabular}{|l|c|}
\hline
\textbf{Reasons} & \textbf{Mentions}\\
\hline
Safest option & 22 \\
The cart shouldn't rely on people getting out of way & 11 \\
No free space for other actions & 9 \\
Bothers no one & 6 \\
Pedestrian might not realize the cart & 6\\
Wait for the left path to get clear & 4
\\
Acting like a pedestrian in the same situation & 3 \\
Pedestrians' right of way (priority) & 3 \\
\hline
\end{tabular}
\end{table}

\begin{figure*}[!b]
\centering
\subfigure[Human-Human interaction]{
\fbox{\includegraphics[height=3.5cm]{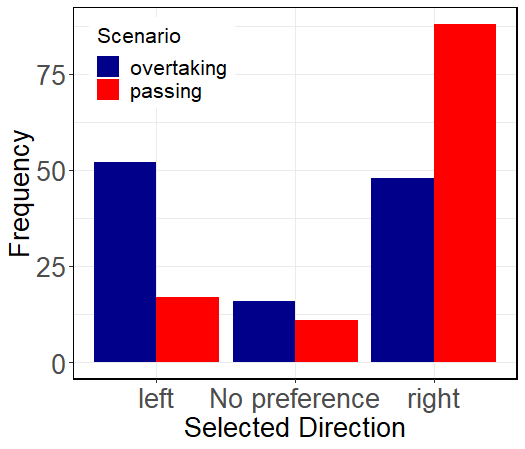}
\Description[A double bar graph for Human-Human interaction scenario]{Figure 2 (a) shows a double bar graph, summarizing the number of times each of the left/right/No-preference direction was chosen by participants for passing and overtaking another person. For passing, right has a higher frequency than left. For overtaking, left has a slightly higher frequency than right. Frequencies for ``no preference" are  lower than the left or right choices for both passing and overtaking choices.}}
\label{fig:HHI}
}\hspace{1cm}
\subfigure[Main Concerns]{
\fbox{\includegraphics[height=3.5cm]{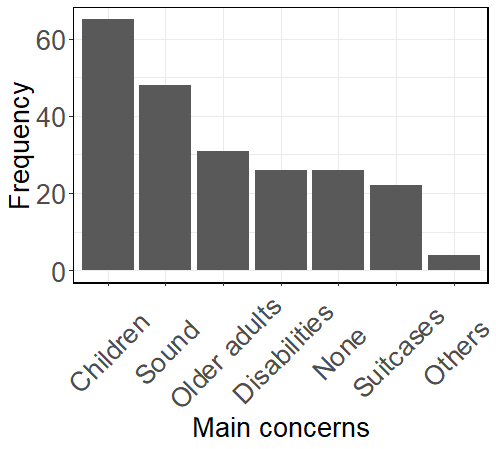}
\Description[A bar graph of the participants’ main concerns]{In Figure 2 (b), the frequency for each of the mentioned concerns is shown. These concerns, from high to low frequencies, are: children, sound, older adults, those with a disability, none, suitcases, and others.}}
\label{fig:MainConc}
}\hspace{1cm}%\hfill
\subfigure[Approach Direction]{
\fbox{\includegraphics[height=3.5cm]{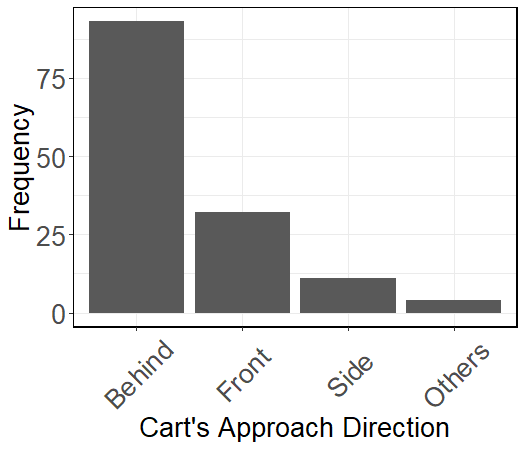}
\Description[A bar graph of the cart’s approach direction that concerns people]{In Figure 2 (c), the frequencies for different approaching directions of the cart are summarized. From high to low frequency, the selected approaches are: behind, front, side, and others.}}
\label{fig:AppDir}
}\hspace{1cm}
\caption{The results of the pre-survey and post-survey questions}
\label{fig:PrePost}
\end{figure*}

\begin{table*}
\centering
\small
\Description[Thematic analysis of participants’ reason in scenario 4]{Four mentioned reasons for preferring the cart to move on the left side of the corridor in scenario 4 are shown. This table also shows three reasons for preferring the right side operation and four reasons for preferring the sides rather than going through the middle.}
\caption{Thematic analysis results on participants' reasons for their preferred operational side of the cart in a corridor in scenario 4}\label{tab:s4_Qual}
\begin{tabular}{|l|l|C{1.5cm}|}
\hline
\textbf{Choice} & \textbf{Reason} & \textbf{Mentions}\\
\hline
\multirow{4}{2.2cm}{The cart moving on left} & Traffic rule: Faster vehicle on the left/overtaking on the left & 37 \\
& Negative feeling about the cart going through the middle & 8 \\
& More intuitive & 4 \\
& Similar to the structure of sidewalk-road & 3 \\
\hline
\multirow{3}{2.2cm}{The cart moving on right} & Traffic rule: the cart should keep on right & 7\\
& Negative feeling about the cart going through the middle & 7 \\
& Safer & 3\\
\hline
\multirow{4}{2.2cm}{Selecting both left and right} & Lower risk of hitting someone  & 3\\
& Safer to have pedestrian on one side & 3 \\
& Better to keep pedestrians and the cart separate & 3\\
& Uncomfortable with the cart in the middle & 2\\
\hline
\end{tabular}
\end{table*}

Figure~\ref{fig:MultChoice} shows the result of the three multiple choice scenarios (scenarios 3, 4, and 7). In scenario 3 where the cart gets blocked behind a single pedestrian, significant differences were found between the preferred choices in both groups ($Pedestrians:{X^2}_{(3, N=59)}=45.34, p<0.001; Passengers:{X^2}_{(3, N=57)}=100.68, p<0.001$). For the pedestrian group, the "clearing the carts' path by moving to the right" behaviour was selected significantly more than "moving to the left" ($p<0.001$) and also "Keep walking straight" ($p<0.001$). In the same scenario, the passenger group selected the behaviour of "staying behind the pedestrian" significantly more than expecting pedestrian A ($p<0.001$) or pedestrians B and C ($p<0.001$) to clear the cart's path.
Based on thematic analysis of reasons, participants in the pedestrian group mainly chose to clear the cart's path as they were uncomfortable with and anxious about the cart's presence behind them and they found it safer for themselves to clear the cart’s path to reduce the risk of getting hit. Not disturbing other pedestrians on the left side that are walking in the opposite direction, was the most mentioned rationale for clearing the cart's path by choosing to move to the right side compared to left. Furthermore, letting faster vehicles overtake on the left was pedestrians' second frequent rationale for the same choice~(Table~\ref{tab:s3_PED_Qual}).
On the other hand, participants in the role of a passenger in the same scenario selected to stay behind the pedestrian because they believed it was the safest option and the cart should not rely on people getting out of its way. They indicated that as there is no free space for taking other actions and pedestrians might not realize the cart, the cart should stay behind the pedestrians until the left path gets clear (see Table~\ref{tab:s3_PAS_Qual}).

In an unidirectional flow of pedestrians in a corridor (scenario 4), the cart moving on the left side of the corridor was preferred significantly more than moving on the right ($p<0.01$) and moving in the middle ($p<0.001$) by the participants in the pedestrian group as shown in Fig. \ref{fig:s4}. In the passenger group, moving in the middle was significantly less preferred as compared with moving on left ($p<0.001$) and moving on the right ($p<0.001$). However, we did not observe any significant difference between the left and right passage of the cart in the passenger group. Results are summarized in Fig. \ref{fig:s4}. 
Obeying the traffic rule of faster vehicle driving on the left was the mostly mentioned reason for preferring cart's left passage (see Table \ref{tab:s4_Qual}). 
The thematic analysis of the reasons also suggested that the participants' concern about the cart passing from the middle was due to the high risk of this option. 

In a bidirectional flow of pedestrians (scenario 7), there was a strong preference in both groups (passengers and pedestrians) for the option where the cart follows the flow of pedestrians walking in the same direction as the cart (group A), compared to all other options of opening a passage through the flow of pedestrians whether from right ($p<0.001$), left ($p<0.001$), or through the middle ($p<0.001$). Results are shown in Fig.~\ref{fig:s7}. No significant difference was found among the other choices (i.e., left, right, and middle passages).

When the cart gets behind an unidirectional flow of pedestrians (scenario 5), the results suggest that requesting a way through is perceived significantly less safe than when the cart just follows the flow of pedestrian ($se=0.16, t=-3.81, p<0.001$).  Participants in the pedestrians group perceived both behaviours less safe than the passengers ($se=0.18, t=-3.07, p<0.01$). 

Further, the passengers found the cart's request for a path through  significantly more dangerous ($p<0.001$) and inappropriate ($p<0.01$), as compared with following the flow of the pedestrians, and rated to be more uncomfortable within that situation ($p<0.05$). Passengers also did not expect the cart to request a way through ($p<0.001$) and their trust level decreased significantly when the cart requested for a path, as compared to when it followed the flow ($p<0.001$). We did not see any significant difference between the preferences of the pedestrians, except that they were not expecting the cart to request a way through compared to staying behind them ($p<0.01$). For all the additional questions, except for the one on awareness, the pedestrian group's ratings were significantly lower than the passenger group's ratings when the cart's behaviour was to follow the crowd.

Comparing the perceived safety of the different behaviours in a ninety-degree approach scenario (scenario 6), the cart’s lateral path change was perceived significantly less safe than yielding to the pedestrian whether by a complete stop ($se=0.11, t=16.36, p<0.001$) or by slowing down ($se=0.11, t=7.53, p<0.001$). Signaling the pedestrian while taking the priority had a significant positive effect on the perceived safety of the cart compared to the same behaviour with no signaling ($se=0.11, t=-6.82, p<0.001$).

The effect of using explicit cues for communicating the intent, compared to implicit cues, was also observed in the turn scenario (scenario 8). Using both flashing lights and a projected path significantly and positively affected the safety ratings ($se=0.14, t=10.84, p<0.001$ and $se=0.14, t=6.4, p<0.001$, respectively). While the passenger group rated flashing lights to be safer than the projected path ($se=0.14, t=4.43, p<0.001$), we did not find a difference in pedestrians group’s perception.

When encountering a child (scenario 9), the cart’s complete stop was perceived not only significantly safer ($se=0.13, t=11.05, p<0.001$), but also more intelligent ($se=0.12, t=12.27, p<0.001$) compared to the behaviour where the cart just adjusts its path to avoid the child. Moreover, passengers had a safer feeling than pedestrians towards the cart’s complete stop in this scenario ($se=0.19, t=2.36, p<0.05$). 

The results of our pre-survey and post-survey questions are shown in Figure~\ref{fig:PrePost}. In human-human interaction, there was a significant difference in the preferred passing direction (${X^2}_{(2, N=116)}=94.88, p<0.001$). Participants had a clear preference for passing another person on the right compared to passing on the left ($p<0.001$). The highest general concerns about the operation of the cart were (a) the presence of children and (b) the cart’s approaching sound not being noticeable, respectively. The results also showed that participants were concerned about the cart approaching from behind more than the front ($p<0.001$) and side ($p<0.001$) approach directions as shown in Fig.\ref{fig:AppDir}.

\section{Discussion}

We conducted an online study to understand people's preferences regarding the behaviour of an autonomous, low speed cart moving among pedestrians. By showing participants sketches of possible interaction scenarios between a passenger cart and pedestrians in an unstructured environment, and providing different options for the cart’s behaviour in each scenario, we assessed people’s expectations of the cart’s behaviour. Participants were randomly assigned to two groups, i,e.  passengers of the cart or pedestrians. We also studied if the traffic rules are expected to be obeyed by the cart despite the unstructured nature of the environment.

While our human-human interaction scenario suggested that participants had clear preferences about the passing direction, we did not observe significant differences between participants' preferences for passing direction in cart-pedestrian interaction (scenario 2). However, when more than one pedestrian was involved,  participants showed a significant tendency to obey traffic rules of passing on the right and overtaking on the left. Speed difference seems to be a major factor governing people’s preferences for the overtaking direction, as the participants mentioned the common rule on the road of higher speed vehicles moving on the left for their reasons~(Table \ref{tab:s4_Qual}). This can also explain the non-significant difference for overtaking direction in the human-human interaction scenario as humans' speeds are not much different from each other. 

Participants' seemed to prefer following traffic rules even in unstructured environments, as shown in our scenarios. For example, the cart taking the left side passage was preferred, which can be similar to the traffic rule that requires faster vehicles to move on the left. Even those who chose the right passage for the cart were referring to a traffic rule that cars generally drive on the right side, viewing the scenario more as a passing situation rather than overtaking.  
Also, the participants in the passenger group appeared to have  a general belief that the cart should not rely on people getting out of its way. Instead, it was perceived the safest for the cart to stay behind the pedestrian/s until the left path gets clear. However, in the pedestrian group, most participants chose to clear the cart’s path despite the presence of obstacles on both sides. Reasons provided by participants showed that participants in the pedestrian group felt uncomfortable about being followed by the cart and they found it safer for themselves to clear the cart’s path to reduce the risk of getting hit.

So, by introducing a passenger-carrying cart even in an unstructured environment, participants seem to still view the scene as a structured environment (e.g., a road) and prefer the cart to follow the traffic rules, which can increase the perceived safety of the cart (Addressing RQ1 and RQ2). 
It is likely that participants have shown this preference as the operation of a motorized vehicle in these unstructured environment is relatively new and the most familiar interactions similar to that are the encounters that happen on the roads. So participants might be carrying over their expectation from the roads to these unstructured environments.

Further, regarding participants' preferences about the behaviour of the cart when negotiating space with pedestrians (RQ3), we found that there is a high tendency to separate the vehicle’s path from the pedestrians' path as much as possible for reducing the risk of injuring someone (scenario 4). This was shown in participant's stated reasons for not selecting the middle passage for the cart in scenario 4~(Table \ref{tab:s4_Qual}). Also, people were mostly expecting the cart to just follow the flow of pedestrians without disturbing them. This was the case for both a unidirectional flow of pedestrian (scenario 5) and a bidirectional flow of pedestrian (scenario 7). The responses to the scenarios which simulated situations where the cart gets blocked behind pedestrian(s) (scenarios 3, 5 and 7) have shed some light on how people think about the priorities in these unstructured environments (Addressing RQ3). The responses showed that people may not necessarily give priority to the vehicle by clearing the path for the cart despite its higher operational speed.

Also, as the presence of children was the mostly mentioned concern people had, and also because children's movements, if moving freely, are highly  unpredictable, it might be better for the cart to do a complete stop  when encountering children (also suggested by scenario 9). The use of some type of alert was also frequently recommended by the participants.

Overall, based on our results we can extract the following tentative norms, which may serve as design suggestions for a navigation algorithm of an autonomous, passenger-carrying cart operating in an unstructured, crowded environment, with the ultimate goal to help increase the perceived safety and acceptability of the cart.

\begin{itemize}[leftmargin=*, noitemsep,topsep=1pt]
    \item Obeying common traffic rules of overtaking on the left and passing on the right when interacting with a crowd.
    \item Keeping the cart's path separate from other pedestrians as much as possible.
    \item Using explicit auditory and visual signal for communicating intent.
    \item Using velocity change instead of lateral deviation of the path for collision avoidance.
    \item Coming to a complete stop instead of bypassing when encountering a child that is running about.
\end{itemize}
As the above mentioned behaviours were perceived safer by the participants in both passenger and pedestrian groups, paying attention to these behaviours could help improve the cart’s acceptance in the crowd.

Our work clearly has multiple limitations. Due to the COVID-19 situation we were not able to conduct any in-person studies with a real golf cart. Our sketched-based scenarios are likely to  have affected the results, as imagining situations based on the sketches would be different from realistic interactions. Also, in those sketches used, it was difficult to provide rich contextual information on the nature of the unstructured environments, the speed and size of the cart, its specific design and communication features (e.g. lights), and most importantly three-dimensional and temporal information which may be crucial in real encounters of passengers  or pedestrians with an autonomous golf car. This could raise doubts on whether the participants had indeed imagined themselves in a truly unstructured environment and also in the correct role by just seeing the top-view of the abstract sketches of the scene, despite clearly pointing to a specific pedestrian in that scene. Future research needs to investigate if the results of this study would translate to real world interactions while also including a wider range of interaction scenarios. It would be also interesting to further study how much the autonomous nature of the cart influenced the responses by comparing it to a human-driven cart in the same scenarios. The type of preferences and social norms studied in this work are also highly cultural dependent and this may limit our results to US and Canada, or countries that follow a right-hand traffic rule. Despite these limitation, this preliminary study could serve as a starting point for understanding people’s expectation of these carts in an unstructured environment, in order to  design autonomous, low-speed vehicles that can successfully operate in crowded areas.

\section{Conclusion}
We studied people’s expectations regarding how an autonomous, low-speed, golf-cart type vehicle should navigate in an unstructured, crowded environment through an online survey. Our results suggested that the cart is not expected to disturb the flow of the pedestrians for getting through the crowd, but, whenever it wants to pass a group of pedestrians, it is expected to follow common traffic rules. While following this norm was considered to be important in the interaction of the cart with a group of pedestrians, we did not find any preference for passing or overtaking direction in a single pedestrian-cart interaction, especially when there was space on both sides of the pedestrian. Results also suggested that pedestrians may mostly react to a cart behind them by clearing its path as they do not not feel comfortable about the cart’s presence behind their backs. This was also consistent with participants' most concerned approach direction of the cart, i.e. being from behind.

While these results are not entirely surprising, and need to be confirmed in future more realistic scenarios and settings, they form a useful starting point for our quest in understanding the norms that people expect a golf cart to follow when operating in unstructured and crowded environments.

\begin{acks}
This research was undertaken, in part, thanks to funding from the Canada 150 Research Chairs Program and NSERC.
\end{acks}

\balance
\bibliographystyle{ACM-Reference-Format}
\bibliography{main}

\newpage
\appendix

\onecolumn
\setcounter{table}{0}
\renewcommand{\thetable}{A\arabic{table}}
\renewcommand*{\arraystretch}{1.8}

\section{Scenarios}

\begin{table} [!h]
    \footnotesize
    \caption{The main scenarios of the survey. Upper-case letters are behaviors for which a separate questionnaire was completed by the participants. The lower-case letters show the multiple-choice questions}\label{tab:scenarios}
\begin{tabular}{|c|c|p{4.5cm}|c|c|p{4.5cm}|}
    \hline \multicolumn{1}{|l|}{No.} & \multicolumn{1}{c|}{Scenario} & \multicolumn{1}{c|}{Behaviours} &
    \multicolumn{1}{|l|}{No.} &
    \multicolumn{1}{c|}{Scenario} & \multicolumn{1}{c|}{Behaviours} \\ \hline

    \noindent 1 &
    \raisebox{-\totalheight}{\includegraphics[width=25mm]{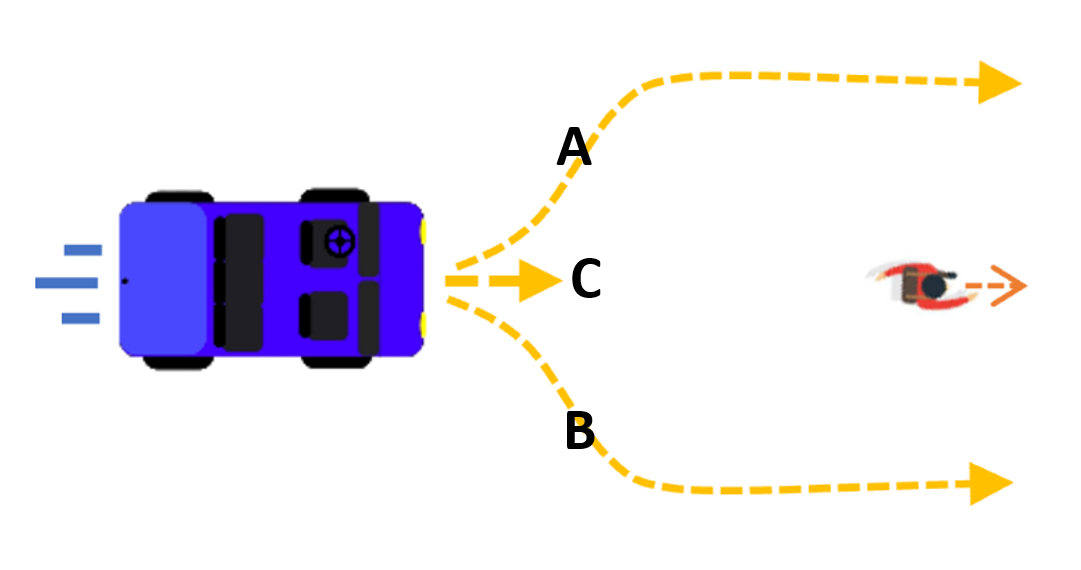}
    \Description[Cart approaching from behind]{A sketched scenario of a cart approaching a single pedestrian from behind with free spaces on both left and right sides of the pedestrian and the cart}} &
    \renewcommand{\theenumi}{\Alph{enumi}}
    \begin{enumerate}[topsep=0pt]
      \item Overtaking on the left.
      \item Overtaking on the right.
      \item Following the pedestrian.
    \end{enumerate} &
    2 & 
    \raisebox{-\totalheight}{\includegraphics[width=25mm]{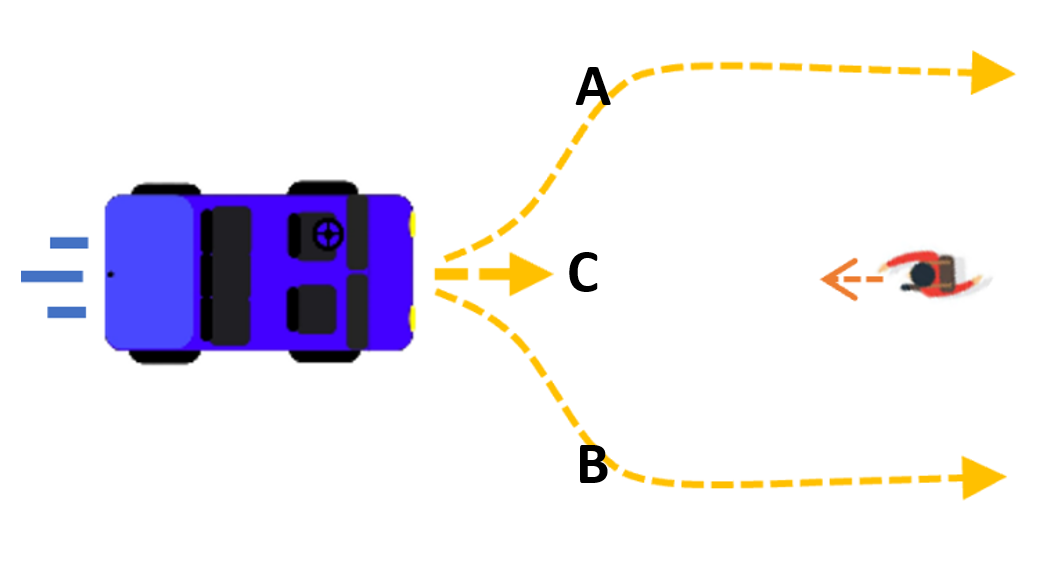}
    \Description[Cart approaching from front]{A sketched scenario of a cart approaching a single pedestrian from front with free spaces on both left and right sides of the pedestrian and the cart }} & 
    \renewcommand{\theenumi}{\Alph{enumi}}
    \begin{enumerate}[topsep=0pt]
      \item Passing on the left.
      \item Passing on the right.
      \item Wait for the pedestrian to clear the path.
    \end{enumerate} 
    \\

    3 & 
    \raisebox{-\totalheight}{\includegraphics[height=30mm]{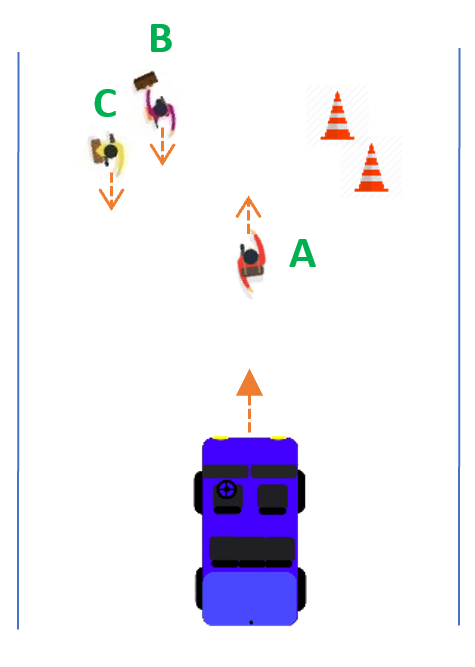}
    \Description[Cart behind a pedestrian with occupied sides]{A sketched scenario of a cart approaching one of the pedestrians (pedestrian A) from behind. On the left side of pedestrian A there are two other pedestrians (B and C) walking in the opposite direction. The right side of A is also occupied by pylons}} & 
    For pedestrians in place of A:
    \renewcommand{\theenumi}{\alph{enumi}}
    \begin{enumerate}[topsep=0pt]
      \item Clearing cart’s path by moving to right.
      \item Clearing cart’s path by moving to left.
      \item Keep walking straight until a place where the cart can overtake me.
      \item Paying no attention 
      \item Others
    \end{enumerate} 
    For passengers:
    \renewcommand{\theenumi}{\alph{enumi}}
    \begin{enumerate}
      \item Expecting A to clear the cart’s path.
      \item Expecting B and C to clear the cart’s path.
      \item Following A until getting to a place where it could overtake.
      \item Others
    \end{enumerate} &
    4 &
    \raisebox{-\totalheight}{\includegraphics[width=25mm]{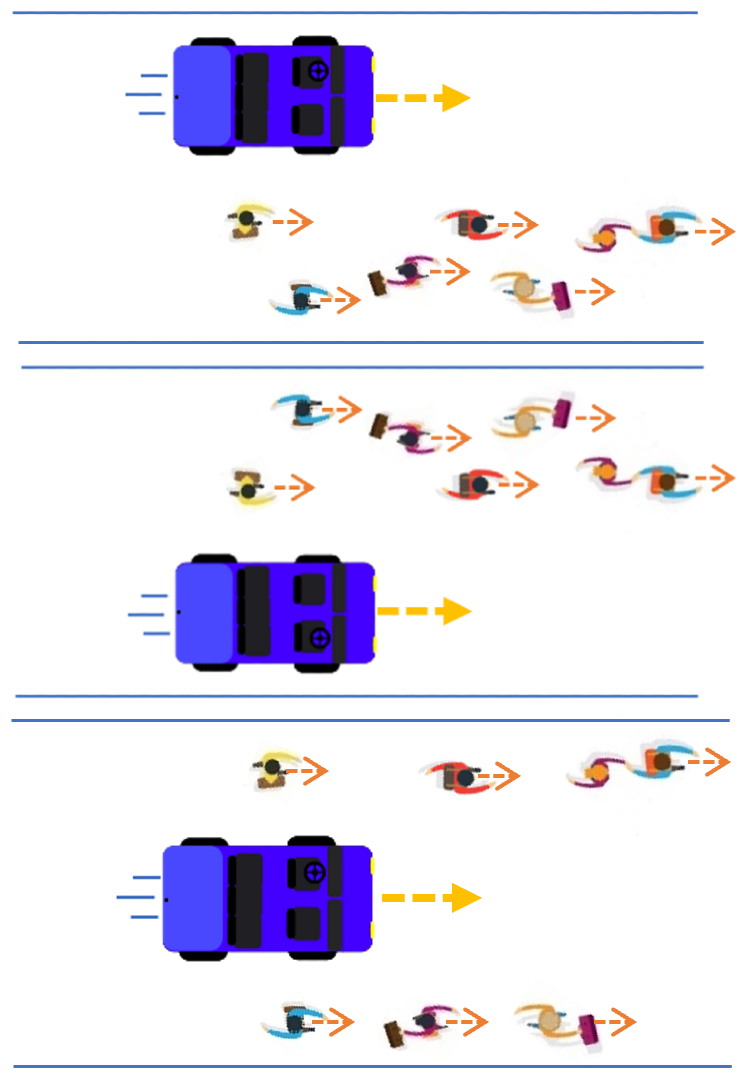}
    \Description[Cart driving in a corridor]{A sketched scenario of a cart driving in a corridor among a unidirectional flow of pedestrians with three options of driving (1) on the left side of the pedestrians, (2) on the right side of the pedestrians, and (3) through the flow with half of the pedestrians on the left of the cart and half of them on the right side of the cart}} &
    \renewcommand{\theenumi}{\alph{enumi}}
    \begin{enumerate}[topsep=0pt]
      \item Moving on the left.
      \item Moving on the right.
      \item Moving in the middle.
      \item No preference.
      \item Others
    \end{enumerate}
    \\
 
    5 & 
    \raisebox{-\totalheight}{\includegraphics[width=30mm]{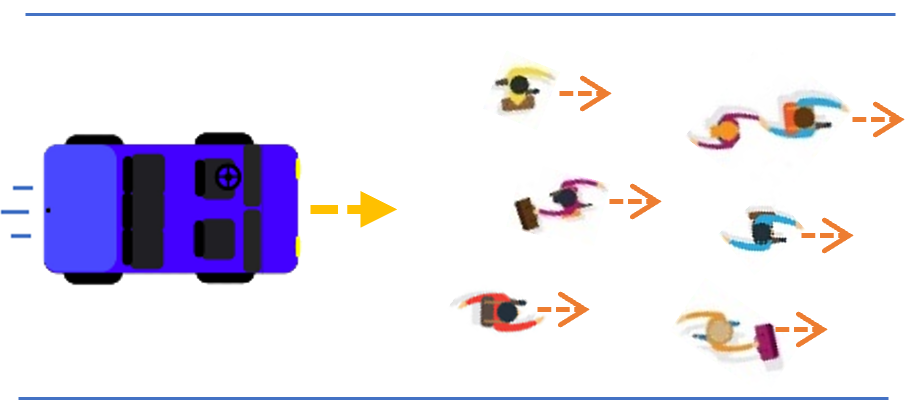}
    \Description[Cart getting blocked in a corridor]{A sketched scenario of a cart getting blocked behind a unidirectional flow of pedestrians in a corridor walking in the same direction as the cart.}} & 
    \renewcommand{\theenumi}{\Alph{enumi}}
    \begin{enumerate}[topsep=0pt]
      \item Following the flow of pedestrians.
      \item Requesting a way through
    \end{enumerate} &

    6 & 
    \raisebox{-\totalheight}{\includegraphics[width=20mm]{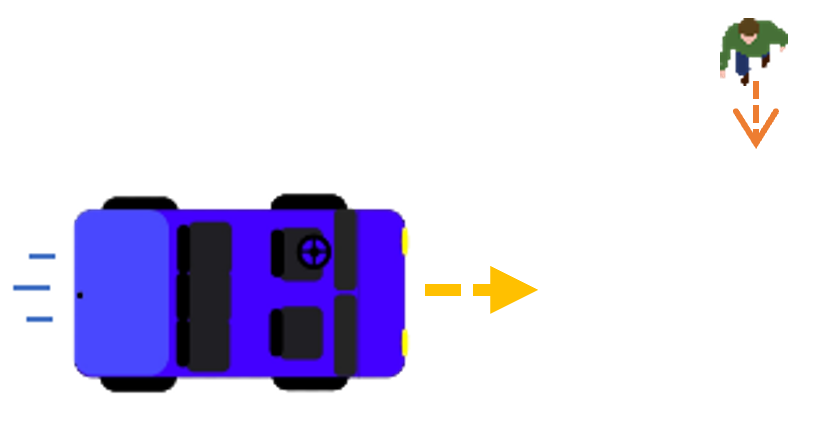}
    \Description[Crossing scenario]{A sketched scenario of a cart and a human approaching in a 90-degree angle in front of the car, in a way that their paths seem to collide}} & 
    \renewcommand{\theenumi}{\Alph{enumi}}
    \begin{enumerate}[topsep=0pt]
      \item Yielding to the pedestrian by stopping.
      \item Yielding to the pedestrian by slowing down.
      \item Adjusting path to the left or right while keeping the same speed.
      \item Taking the priority to cross by just continuing the path without decelerating.
      \item Using visual and auditory signal for showing the intention to cross first and taking the priority to cross.
    \end{enumerate} 
    \\
    
    7 &
    \raisebox{-\totalheight}{\includegraphics[height=40mm]{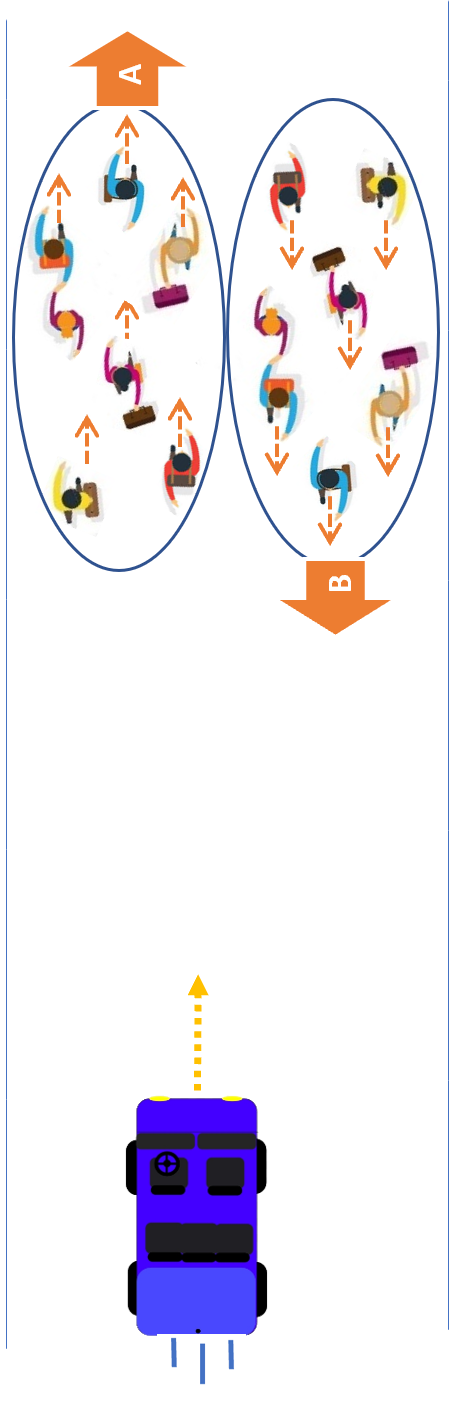}
    \Description[Cart in a bidirectional flow of pedestrian]{A sketched scenario of a cart getting into a corridor with two groups of pedestrians (A and B) walking in the opposite directions. Group A is walking in the same direction as the cart, on the left of the corridor, and group B is walking in the opposite direction on the right side of the corridor}} & 
    \renewcommand{\theenumi}{\alph{enumi}}
    \begin{enumerate}[topsep=0pt]
      \item Following group A.
      \item Opening a passage in the middle.
      \item Opening a passage on the left side of the corridor.
      \item Opening a passage on the right side of the corridor.
      \item Others
    \end{enumerate} &
   
    8 & 
    \raisebox{-\totalheight}{\includegraphics[width=25mm]{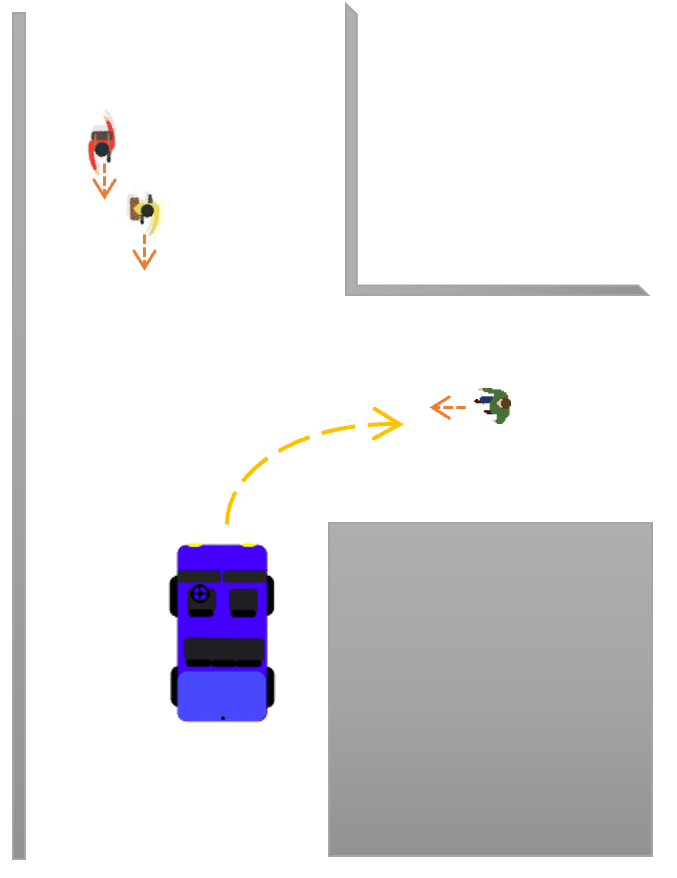}
    \Description[Turn scenario]{A sketched scenario of a cart getting into a three-way-like intersection. The cart wants to turn to the right (shown through a line showing the expected path of the cart) where a pedestrian is, and that pedestrian is also walking towards the intersection. There are two other pedestrians on the third side of the intersection moving towards the cart}} & 
    \renewcommand{\theenumi}{\Alph{enumi}}
    \begin{enumerate}[topsep=0pt]
      \item Showing turn intention implicitly only through movement.
      \item Showing turn intention using flashing lights.
      \item Showing turn intention by projecting the future path on the floor.
    \end{enumerate} 
    \\ %\cline{4-6}
    9 & 
    \raisebox{-\totalheight}{\includegraphics[width=25mm]{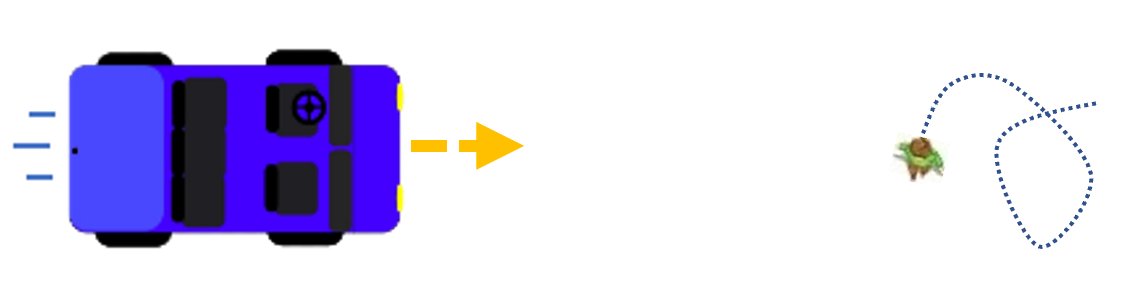}
    \Description[Cart encountering a child]{A sketched scenario of a cart approaching a child on its straight path while the child is roaming around (shown by lines around the child)}} & 
    \renewcommand{\theenumi}{\Alph{enumi}}
    \begin{enumerate}[topsep=0pt]
      \item Stopping and waiting for the child to get into a safe zone.
      \item Decelerating and changing path to the right or left for avoiding the child.
    \end{enumerate} &
        Pre & 
    \raisebox{-\totalheight}{\includegraphics[height=15mm]{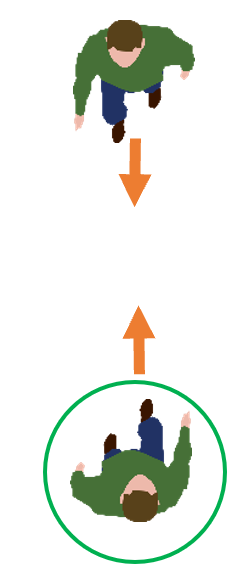}
    \Description[Human-Human interaction]{A sketched scenario of two humans approaching each other face to face}} & 
    \renewcommand{\theenumi}{\alph{enumi}}
    \begin{enumerate}[topsep=0pt]
      \item Passing on the left.
      \item Passing on the right.
      \item No preferred direction.
    \end{enumerate} \\
   \hline
\end{tabular}      
\end{table}
%}

\end{document}